\documentclass[
aps,
prd,
onecolumn,
10pt,
superscriptaddress,
floatfix,
longbibliography,
nofootinbib,
preprintnumbers
]{revtex4-2}
\usepackage{graphicx}
\usepackage{dcolumn}
\usepackage{bm}
\usepackage{ulem}
\usepackage{amsmath}
\usepackage{amssymb}
\usepackage{mathrsfs}
\usepackage{txfonts}
\usepackage{hyperref}
\usepackage{color} 
\usepackage{xcolor}
\usepackage{listings}
\usepackage{cprotect}

\usepackage{tikz}
\usetikzlibrary{quantikz2}

\newcommand{\Tr}{{\rm Tr}}
\newcommand{\e}{{\rm e}}

\newcommand{\im}{\mathrm{i}}

\hypersetup{
  colorlinks=true,
  linkcolor=[rgb]{0.60,0.00,0.00},
  citecolor=[rgb]{0.00,0.00,0.60},
  urlcolor=[rgb]{0.00,0.00,0.60},
  setpagesize=false
}

\definecolor{codegreen}{rgb}{0,0.6,0}
\definecolor{codegray}{rgb}{0.5,0.5,0.5}
\definecolor{codepurple}{rgb}{0.58,0,0.82}
\definecolor{backcolour}{rgb}{0.95,0.95,0.92}

\lstdefinestyle{mystyle}{
    backgroundcolor=\color{backcolour},   
    commentstyle=\color{codegreen},
    keywordstyle=\color{magenta},
    numberstyle=\tiny\color{codegray},
    stringstyle=\color{codepurple},
    basicstyle=\ttfamily\footnotesize,
    breakatwhitespace=false,         
    breaklines=true,                 
    captionpos=b,                    
    keepspaces=true,                 
    numbers=left,                    
    numbersep=5pt,                  
    showspaces=false,                
    showstringspaces=false,
    showtabs=false,                  
    tabsize=2
}

\begin{document}

\title{Floquet prethermalization of \texorpdfstring{${\bf Z}_2$}{Z2} lattice gauge theory on superconducting qubits}
\author{Tomoya Hayata}
\email{hayata@keio.jp}
\affiliation{Departments of Physics, Keio University School of Medicine, 4-1-1 Hiyoshi, Kanagawa 223-8521, Japan}
\affiliation{Interdisciplinary Theoretical and Mathematical Sciences Program (iTHEMS), RIKEN, Wako, Saitama 351-0198, Japan}

\author{Kazuhiro~Seki}
\email{kazuhiro.seki@riken.jp}
\affiliation{Quantum Computational Science Research Team, RIKEN Center for Quantum Computing (RQC), Saitama 351-0198, Japan}

\author{Arata Yamamoto}
\email{arayamamoto@nt.phys.s.u-tokyo.ac.jp}
\affiliation{Department of Physics, The University of Tokyo, Tokyo 113-0033, Japan}

\preprint{RIKEN-iTHEMS-Report-24}

\begin{abstract}
Simulating nonequilibirum dynamics of a quantum many-body system is one of the promising applications of quantum computing. We simulate the time evolution of one-dimensional ${\bf Z}_2$ lattice gauge theory on IBM's superconducting 156-qubit device ibm\_fez. We consider the Floquet circuit made of the Trotter decomposition of Hamiltonian evolution and focus on its dynamics toward thermalization. Quantum simulation with the help of error mitigation is successful in running the Floquet circuit made of $38$ and $116$ qubits up to $10$ Trotter steps in the best case. This is enough to reach the early stage of prethermalization. Our work would be a benchmark for the potential power of quantum computing for high-energy physics problems.

\end{abstract}

\maketitle

\section{Introduction}

Simulating non-equilibrium quantum systems is one of the promising targets of quantum computers.
Non-equilibrium dynamics of quantum field theory will be able to be studied by quantum simulation, but we may need to await the realization of fault-tolerant quantum computers. At present, the depth of executable circuits is severely limited due to harmful device noise. This implies that the number of Trotter steps is severely limited, and we can do simulations only up to very short times to make Trotter errors under control.
The practical implementation of continuous time evolution, say, Hamiltonian simulation, with classically intractable system size may be unrealistic for near-term quantum devices. Discrete time evolution, such as Floquet circuits~\cite{Eckstein:2023sjk,Yang:2023nak,Shinjo:2024vci,Seki:2024rfx}, is a first step toward practical application.

Making time step very large in the Lie-Suzuki-Trotter decomposition of Hamiltonian evolution, we obtain a quantum circuit feasible for simulating with near-term quantum devices. Although Trotter error is no longer bounded when the time step exceeds a threshold~\cite{Heyl2019}, and we cannot faithfully simulate Hamiltonian evolution, such a quantum circuit exhibits interesting non-equilibrium dynamics as Floquet system. It is known that Floquet system eventually heats up to infinite temperature, by acquiring energy continuously from driving force, but the heating time can be exponentially long with driving frequency i.e., inverse of time step~\cite{Lazarides2014,DAlessio2014,Abanin2015,Mori2016,Kuwahara2016,Mori:2017qhg}. Indeed, Floquet system prethermalizes before reaching to infinite temperature state. We refer to the phenomena as Floquet prethermalization. It was proposed to study the prethermal state in classically intractable system using near-term quantum devices~\cite{Yang:2023nak}. 

In this paper, we carry out the digital quantum simulations of Floquet circuits in lattice gauge theory.
We consider the simplest model of lattice gauge theory; the ${\bf Z}_2$ lattice gauge theory with fermions on a one-dimensional chain with open boundaries.
In this geometry, all the gauge fields can be eliminated by solving the Gauss law constraints, and the local gauge-fermion interaction is replaced by the non-local fermion-fermion interaction, i.e., the long-range Coulomb interaction. While such a reduction is used in many benchmark studies of quantum simulations~\cite{Klco:2018kyo,deJong:2021wsd,Nguyen:2021hyk,Atas:2022dqm,Schuster:2023klj,Angelides:2023noe,Farrell:2023fgd,Farrell:2024fit}, it cannot be generalized to higher-dimensional gauge theories. The simulation in the presence of fermions and gauge fields is a good exercise toward the future~\cite{Mildenberger:2022jqr,Pomarico:2023png,Charles:2023zbl,Hayata:2023skf}. In this study we do not remove fermions or gauge fields but simulate both of them.
From another perspective, all to all interactions use $\mathcal{O}(N^2)$ two-qubit gates in one Trotter step (or more if we have only local connectivity) with the system size $N$, which are too many, and must be approximated on near-term quantum devices~\cite{Farrell:2024fit}.

We used IBM's latest device ``ibm\_fez'', which consists of 156 superconducting qubits, for quantum simulations.
Error mitigation is necessary for interpreting the data by noisy simulation.
We examined three mitigation schemes.
The first one is zero noise extrapolation (ZNE) available from Qiskit's estimator~\cite{qiskit_paper}.
The second and third ones are mitigation of amplitude damping by division. 
In the second method, the damping rate is estimated by measuring the Gauss law operator. 
In the third method, the damping rate is estimated by simulating a Floquet circuit with special parameters such that the circuit becomes trivial if there exist no noises~\cite{Shinjo:2024vci}. 

The rest of this paper is organized as follows. We introduce the ${\bf Z}_2$ lattice gauge theory and the Floquet circuit in Sec.~\ref{sec:methods}. 
We first show the classical simulations with the matrix product state (MPS) ansatz in Sec.~\ref{sec:classical}. 
The parameter dependencies of Floquet prethermalization are checked by the classical MPS simulations. 
After that, we show the main results of quantum simulations in Sec.~\ref{sec:quantum}. We explain circuit details, show the results with three mitigation methods, and discuss the interpretation of the results based on the effective fidelity and circuit volume of the simulated circuits. 
Finally, we give some comments on future applications of quantum computers to the high-energy physics in Sec.~\ref{sec:summary}.

\section{Methods}
\label{sec:methods}
\subsection{\texorpdfstring{${\bf Z}_2$}{Z2} lattice gauge theory}

Here we overview the Hamiltonian formulation of the $(1+1)$-dimensional ${\bf Z}_2$ lattice gauge theory with the Wilson fermion.
We consider a one-dimensional lattice with open boundary conditions, which has $N$ sites and $(N-1)$ links.
The total state vector can be written as
\begin{equation}
 |\Psi\rangle = \prod_{x=1}^{N-1} |g(x,x+1)\rangle \prod_{x=1}^{N} |\psi_1(x)\rangle |\psi_2(x)\rangle ,
\end{equation}
where $x$ labels site position.
The gauge fields are defined on links and each field $|g(x,x+1)\rangle$ at the link in between $x$ and $x+1$ is two dimensional.
The link operator $Z_g(x,x+1)$ and the conjugate operator $X_g(x,x+1)$ act on it.
Fermions are two-component spinors, and each component $|\psi_i(x)\rangle$ is two-dimensional.
The creation and annihilation operators satisfy the anti-commutation relation $\{\psi_i(x),\psi^\dagger_j(y)\}=\delta_{xy}\delta_{ij}$.
The Hamiltonian is given by the gauge part
\begin{equation}
 H_g = - K \sum_{x=1}^{N-1} X_g(x,x+1) ,
\end{equation}
the fermion part
\begin{equation}
H_f = \sum_{x=1}^{N} (1+m) \psi^\dagger(x) \gamma^0 \psi(x) ,
\end{equation}
and the gauge-fermion coupling part
\begin{equation}
\begin{split}
H_{gf} = &-\frac{1}{2} \sum_{x=1}^{N-1} Z_g(x,x+1) \big\{ \psi^\dagger(x) \gamma^0 (1-\gamma^1) \psi(x+1) 
 \\
&+ \psi^\dagger(x+1) \gamma^0 (1+\gamma^1) \psi(x)\big\}  ,
\end{split}
\end{equation}
where $K$ and $m$ are the gauge parameter and fermion mass, respectively.
We used the lattice unit and eliminated the lattice spacing.
Taking $\gamma^0=\sigma_1$, $\gamma^1=\sigma_3$, and performing the Jordan-Wigner transformation with open boundary conditions, we can rewrite the Hamiltonian terms as
\begin{equation}
 H_f = \frac12 \sum_{x=1}^N (1+m) \left\{ X_1(x) X_2(x) + Y_1(x) Y_2(x) \right\}
\label{eqHW_f1}
\end{equation}
and
\begin{equation}
 H_{gf} = -\frac12 \sum_{x=1}^{N-1} Z_g(x,x+1) \left\{ X_1(x) X_2(x+1) + Y_1(x) Y_2(x+1) \right\} .
\label{eqHW_f2}
\end{equation}
The theory can be mapped to $(3N-1)$-qubit systems with linear connectivity~\cite{Zache:2018jbt}.
(If we employ the staggered fermion,  we need $4N-1$ qubits to simulate the same physical volume because of the gauge interaction between $\psi_1(x)$ and $\psi_2(x)$.)

The theory has local ${\bf Z}_2$ gauge symmetry. We consider the theory in the presence of external static charges like infinitely heavy electrons.
The Gauss law operator is defined by
\begin{equation}
\begin{split}
 G(x) &= X_g(x-1,x) e^{-i\pi (\rho(x)+\rho_{\rm s}(x))} X_g(x,x+1) \\
 &= -X_g(x-1,x) Z_1(x)Z_2(x) e^{-i\pi \rho_{\rm s}(x)} X_g(x,x+1)
\end{split}
\end{equation}
with the dynamical charge density
\begin{equation}
\begin{split}
 \rho(x) &= \psi^\dagger_1(x) \psi_1(x) + \psi^\dagger_2(x) \psi_2(x)-1 \\
 &= \frac12 \{ Z_1(x) + Z_2(x) \} ,
\end{split}
\end{equation}
and the external charge density $\rho_{\rm s}(x)$. We take
\begin{align}
\label{eq:external_charges}
  \rho_{\rm s} =
  \left\{
    \begin{array}{lll}
      \displaystyle{1,}
      & \displaystyle{x=1,N},
      \\
      \displaystyle{0,}
      & {\rm otherwise},
    \end{array}
  \right.
\end{align}
i.e., we put charges at the boundaries of the system as external probes.
The physical state satisfies the Gauss law constraint
\begin{equation}
G(x)|\Psi\rangle=|\Psi\rangle
\end{equation}
for all $x$. Here, the operators outside the boundaries are understood as $X_g(-1,0)=X_g(N,N+1)=1$.

\subsection{Floquet Circuit}

We simulate the Floquet circuit obtained from the first-order Lie-Suzuki-Trotter decomposition of Hamiltonian evolution.
The Floquet evolution operator over a single period reads
\begin{equation}
\label{eq:U_F}
    U_\mathrm{F}
    =
    \e^{-\im H_{gf} dt} \e^{-\im (H_{f}+H_{g}) dt} ,
\end{equation}
where $dt$ is the step size.
The time evolution with $U_\mathrm{F}$ can be understood as
a periodically driven quantum system described by the time-dependent Hamiltonian
\begin{align}
\label{eq:periodic_H}
  H(t) =
  \left\{
    \begin{array}{lll}
      \displaystyle{H_{f}+H_{g},}
      & \displaystyle{t\in [0, T/2)},
      \\
      \displaystyle{H_{gf},}
      & \displaystyle{t\in[T/2,T)}
    \end{array}
  \right.
\end{align}
with the driving period $T=2dt$.
The evolution of the state vector over $N_t$ Trotter steps is given by
\begin{equation}
    |\Psi(N_t)\rangle = (U_\mathrm{F})^{N_t} |\Psi(0)\rangle .
\end{equation}
We choose the groundstate of $H_{f}+H_{g}$ in the presence of external charges~\eqref{eq:external_charges} as the initial state $|\Psi(0)\rangle$, which is easily prepared on circuits since $[H_{f},H_{g}]=0$. The fermion part of $|\Psi(0)\rangle$ is given by a product of Bell pairs, while the gauge field part of $|\Psi(0)\rangle$ is given by a product of eigenstates of operators $X_g(x,x+1)$. The evolution conserves the Gauss law constraint because $[U_\mathrm{F},G(x)]=0$.
To study thermalization in the Floquet circuit, we compute the expectation value of the local fermion Hamiltonian
\begin{equation}
\mathscr{H}(x_c) = \frac12 (1+m) \{ X_1(x_c) X_2(x_c) + Y_1(x_c) Y_2(x_c) \}
\label{eqhc}
\end{equation}
and the Gauss law operator
\begin{equation}
 G(x_c) = -X_g(x_c-1,x_c) Z_1(x_c)Z_2(x_c) X_g(x_c,x_c+1)
 \label{eqgc}
\end{equation}
at the center $x_c=(N+1)/2$.
We fix $K=m=1.0$ for the simulations.

\section{Classical MPS simulation}
\label{sec:classical}

\begin{figure}[t]
  \centering
  \includegraphics[width=.48\textwidth]
  {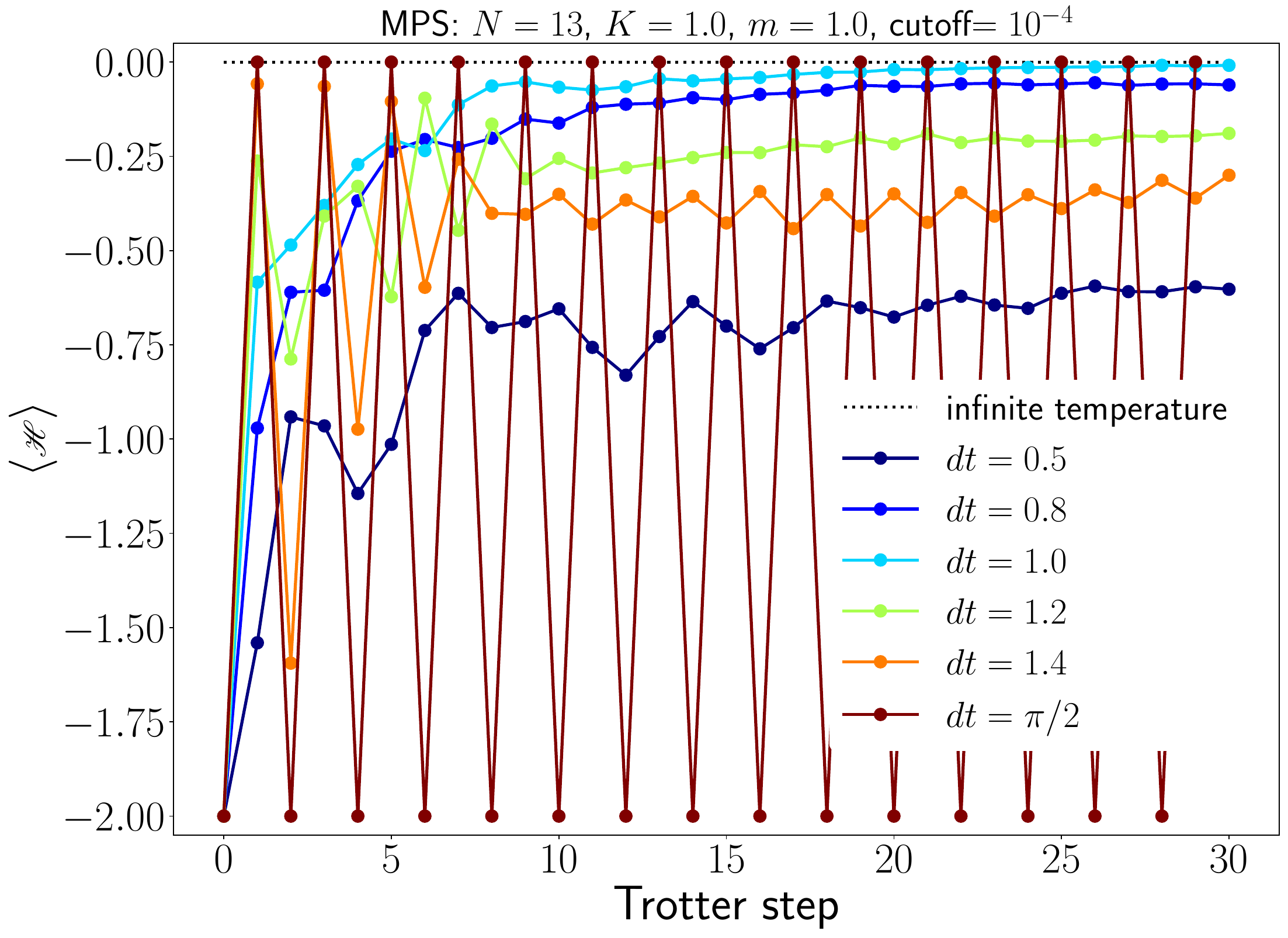}
  \caption{
  \label{fig:mps_tau_dependence}
   Classical simulation results of Trotter-time step dependence of Floquet prethermalization. The system size is $N=13$. The gauge parameter is $K=1.0$, and the fermion mass is $m=1.0$. The cutoff for tensor contraction is $10^{-4}$, which is small enough for the convergence of results. Lines are just for eye guides.
  }
\end{figure}

Before discussing quantum simulation results using IBM's superconducting quantum devices, let us first show classical simulation results based on MPSs. MPSs are tensor networks that are particularly useful for describing a one-dimensional gapped quantum system including our ${\bf Z}_2$ lattice gauge theory. The purpose of the classical simulations is to estimate the system-size and step-size dependence of Floquet circuits. We used iTensor~\cite{itensor} for the classical MPS simulations. For another application of iTensor to lattice gauge theories, see, e.g.,~\cite{Hayata:2023pkw}.

First, we fix the system size with $N=13$ (as seen below, $N=13$ is large enough to discuss the large volume limit).
We show the dependence on the Trotter step size $dt$ in Fig.~\ref{fig:mps_tau_dependence}.
As a representative of small $dt$, which is necessary for Hamiltonian evolution, we show the Trotter evolution with $dt=0.5$.
Thermalization is reached after many Trotter steps ($>25$-$30$ steps).
It is too hard of a task for present noisy quantum devices.
To find more rapid thermalization that would be simulatable even in present noisy devices, we make the step size very large.
We see ``prethermalization'', i.e., plateau behavior before true thermalization to infinite temperature states, without oscillation for $dt=1.0$, and with oscillation for $dt=1.4$. 
There are prethermalization plateaus during $8$-$16$ Trotter steps for $dt=1.0$, and $8$-$25$ Trotter steps for $dt=1.4$. 
Those Floquet circuits may reach infinite temperature state after prethermalization plateau as we can see in the case of $dt=1.0$ (see Fig.~\ref{fig:mps_tau_dependence}), but this needs much more Trotter steps. We employ $dt=1.0$ and $dt=1.4$, and focus on the prethermal plateau regimes in the following quantum simulation with real devices.

\begin{figure}[t]
  \centering
  \includegraphics[width=.48\textwidth]
  {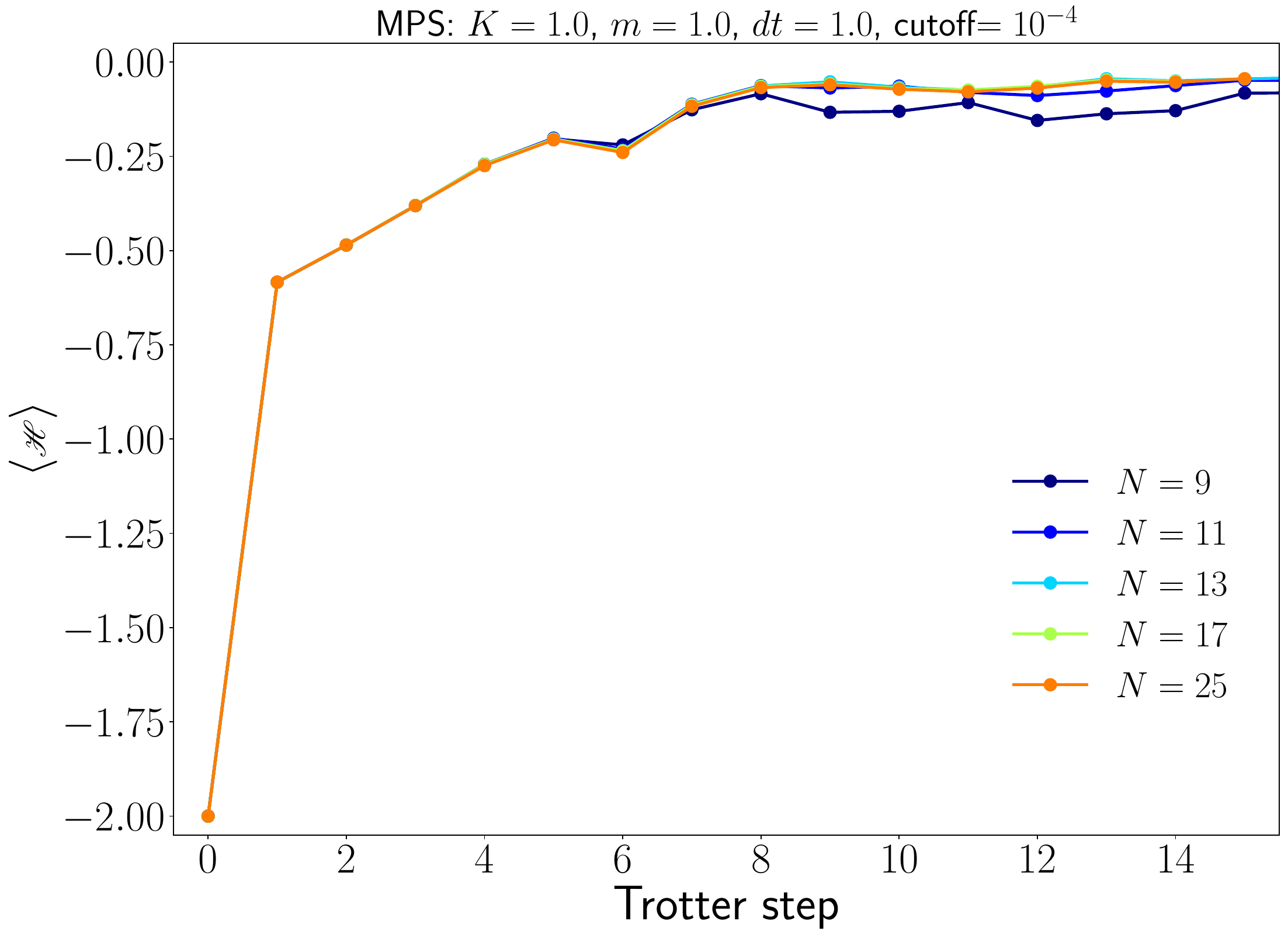}
  \caption{
  \label{fig:mps_N_dependence_1.0}
   Classical simulation results of system-size dependence of Floquet prethermalization. The step size, gauge parameter, and fermion mass are $dt=1.0$, $K=1.0$, and $m=1.0$, respectively. The cutoff for tensor contraction is $10^{-4}$, which is small enough for the convergence of results. Lines are just for eye guides.
  }
\end{figure}
\begin{figure}[t]
\centering
  \includegraphics[width=.48\textwidth]
  {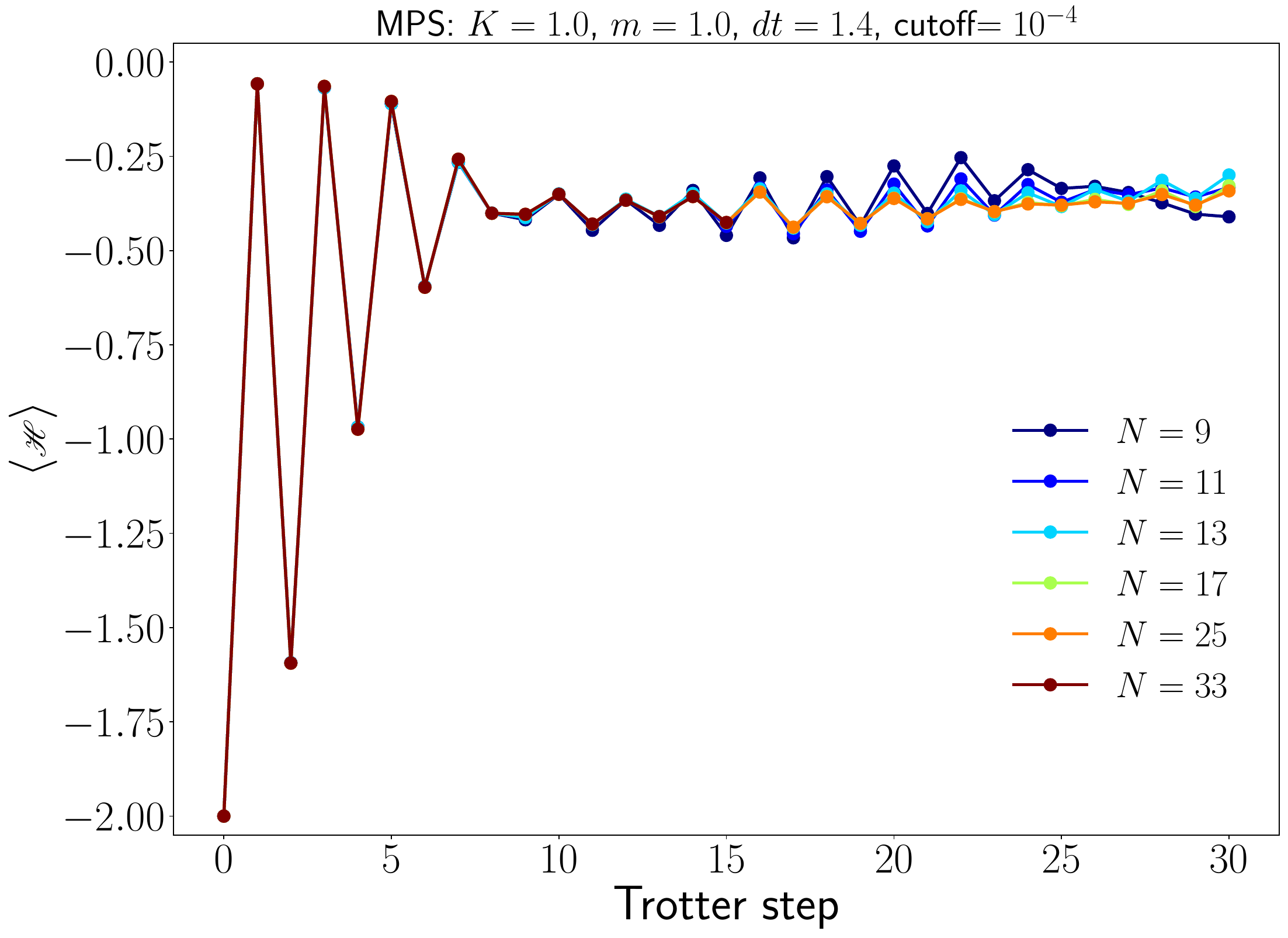}
  \caption{
  \label{fig:mps_N_dependence_1.4}
   Classical simulation results of system-size dependence of Floquet prethermalization. The step size, gauge parameter, and fermion mass are $dt=1.0$, $K=1.0$, and $m=1.0$, respectively. The cutoff for tensor contraction is $10^{-4}$, which is small enough for the convergence of results.  Lines are just for eye guides.
  }
\end{figure}

We show the system size dependence in Figs.~\ref{fig:mps_N_dependence_1.0} and~\ref{fig:mps_N_dependence_1.4}. In both cases, $N=13$ is large enough to simulate the prethermalization dynamics, namely Floquet circuits with up to $15$ or $20$ Trotter steps. Importantly, this is much smaller than the number of qubits inside causal cones at $15$ or $20$ Trotter steps in a large system. We note that the state at late times with $dt=1.0$ is closer to infinite temperature state than $dt=1.4$, and computationally demanding in MPS calculations. This is the reason why we simulate only up to $15$ Trotter steps for $dt=1.0$ with large system sizes. We note also that true thermalization to infinite temperature states shows more severe system-size dependence as seen in Fig.~\ref{fig:mps_N_dependence_1.4}, although it may not be distinguishable in noisy quantum simulation.

\section{Quantum simulation}
\label{sec:quantum}

\subsection{Qiskit compiling details}

We performed quantum simulation using IBM's latest superconducting quantum device named ibm\_fez. The device consists of $156$ qubits with heavy-hex connectivity. 
Mitigation of errors is inevitable for present noisy quantum devices. We tested three mitigation methods using $N=13$ ($38$ qubits), and then apply the best one for simulating a larger system, $N=39$ ($116$ qubits), to see a potential power of present quantum devices.

As for gate counting of the Floquet circuit, since $e^{-i\frac{\theta}{2} (XX+YY)}$ uses two CZ gates~\cite{PhysRevA.69.010301,Foss-Feig:2020pke}:
\begin{widetext}
\begin{equation}
e^{-i\frac{\theta}{2} (XX+YY)}=
   \begin{quantikz}[thin lines]
\lstick{$|\psi_1(x)\rangle$}    & \gate{R_X(\frac{\pi}{2})}  & \ctrl{1} &\gate{R_X(\theta)} 
  & \ctrl{1}  & \gate{R_X(-\frac{\pi}{2})} &  \\
\lstick{$|\psi_2(x)\rangle$}    & \gate{R_X(\frac{\pi}{2})} & \targ{} & \gate{R_z(\theta)}
  & \targ{} & \gate{R_X(-\frac{\pi}{2})} & 
\end{quantikz} ,
\end{equation}
\end{widetext}
$\e^{-\im (H_{f}+H_{g}) dt}$ uses $2N$ CZ gates for each Trotter step.
We follow Ref.~\cite{Charles:2023zbl} for the circuit implementation of $\e^{-\im H_{gf} dt}$. 
We need six CZ gates to implement $e^{i\theta(XZX+YZY)}$:
\begin{widetext}
\begin{equation}
\begin{split}
&e^{i\theta (XZX+YZY)}=
\\
&    \begin{quantikz}[thin lines]
 \lstick{$|\psi_2(x)\rangle$}  & & \targ{} &  & \gate{H} & \targ{}  & \gate{H} & \gate{S} &  &  & & \targ{} & \gate{S^\dag} & \\
  \lstick{$|g(x,x+1)\rangle$} & \gate{H} & \ctrl{-1} & \ctrl{1}   & \gate{R_X(\theta)}  & \ctrl{-1} & \gate{Z} & \ctrl{1} & \gate{R_X(\theta)} &  & \ctrl{1} & \ctrl{-1} & \gate{H} & \\
  \lstick{$|\psi_1(x+1)\rangle$} & & & \targ{} & \gate{H} & & & \targ{} & \gate{H} & \gate{S} &  \targ{} & & \gate{S^\dag} &
\end{quantikz} ,
\end{split}
\end{equation}
\end{widetext}
so that $\e^{-\im H_{gf} dt}$ uses $6(N-1)$ CZ gates for each Trotter step. 
Furthermore, the initial state is prepared on the circuit as
\begin{equation}
\begin{split}
&|\Psi(0)\rangle=
\\
&    \begin{quantikz}[thin lines]
 \lstick{$|0_1(1)\rangle$} & \gate{H} & \gate{Z} & \ctrl{1} & \\
 \lstick{$|0_2(1)\rangle$} &  & \gate{X} & \targ{} & \\
 \lstick{$|0_g(1,2)\rangle$} & \gate{H} & \gate{Z} & & \\
   &  \wireoverride{n}  \\
   &  \wireoverride{n} & \lstick{$\vdots$} \wireoverride{n}  \\
   &  \wireoverride{n}  \\
 \lstick{$|0_g(N-1,N)\rangle$} & \gate{H} & \gate{Z} & & \\
 \lstick{$|0_1(N)\rangle$} & \gate{H} & \gate{Z} & \ctrl{1} & \\
 \lstick{$|0_2(N)\rangle$} &  & \gate{X} & \targ{} &  
\end{quantikz} .
\end{split}
\end{equation}
In short, the total number of CZ gates used for $N_t$ Trotter steps of the $N$-site system is thus $(8N-6)N_t+N$.

We used Qiskit's estimator to compute the expectation values of $\mathscr{H}(x_c)$~\eqref{eqhc} and $G(x_c)$~\eqref{eqgc}. 
The total number of shots used to compute each expectation value is $100000$. 
We enabled readout error mitigation based on the Pauli twirling, named as twirled readout error eXtinction (TREX), in all the experimental data. 
The dynamical decoupling was disabled in all the simulations.
We enabled Pauli twirling of two-qubit gates.
The number of twirled circuits (\texttt{num\_randomizations} in Qiskit's twirling options) was set to $16$ in the simulations for $N=13$ with ZNE, $64$ in the simulations for $N=13$ with other mitigation, and 32 in the simulations for $N=39$. 
We used default options for ZNE.
Changing \texttt{num\_randomizations} in $16,32,64,96$ for $N=13$, we found that $32$ or $64$ was optimal for our circuit.
However, we could not use the optimal value with ZNE due to the limitation of computational resource.

In addition, it may be optional since the growth of the effective causal cone is much slower than that of the causal cone, but we remove unnecessary gates outside the causal cone using pytket~\cite{tket_paper}.
We first wrote the Qiskit code of a circuit and translated it to tket. Then, we removed unnecessary gates using tket's compilation pass \texttt{RemoveDiscarded}, and re-translated the reduced circuit to Qiskit. Finally, we executed the built Qiskit code on the real device.

\subsection{zero noise extrapolation}
\begin{figure}[t]
  \centering
  \includegraphics[width=.48\textwidth]
  {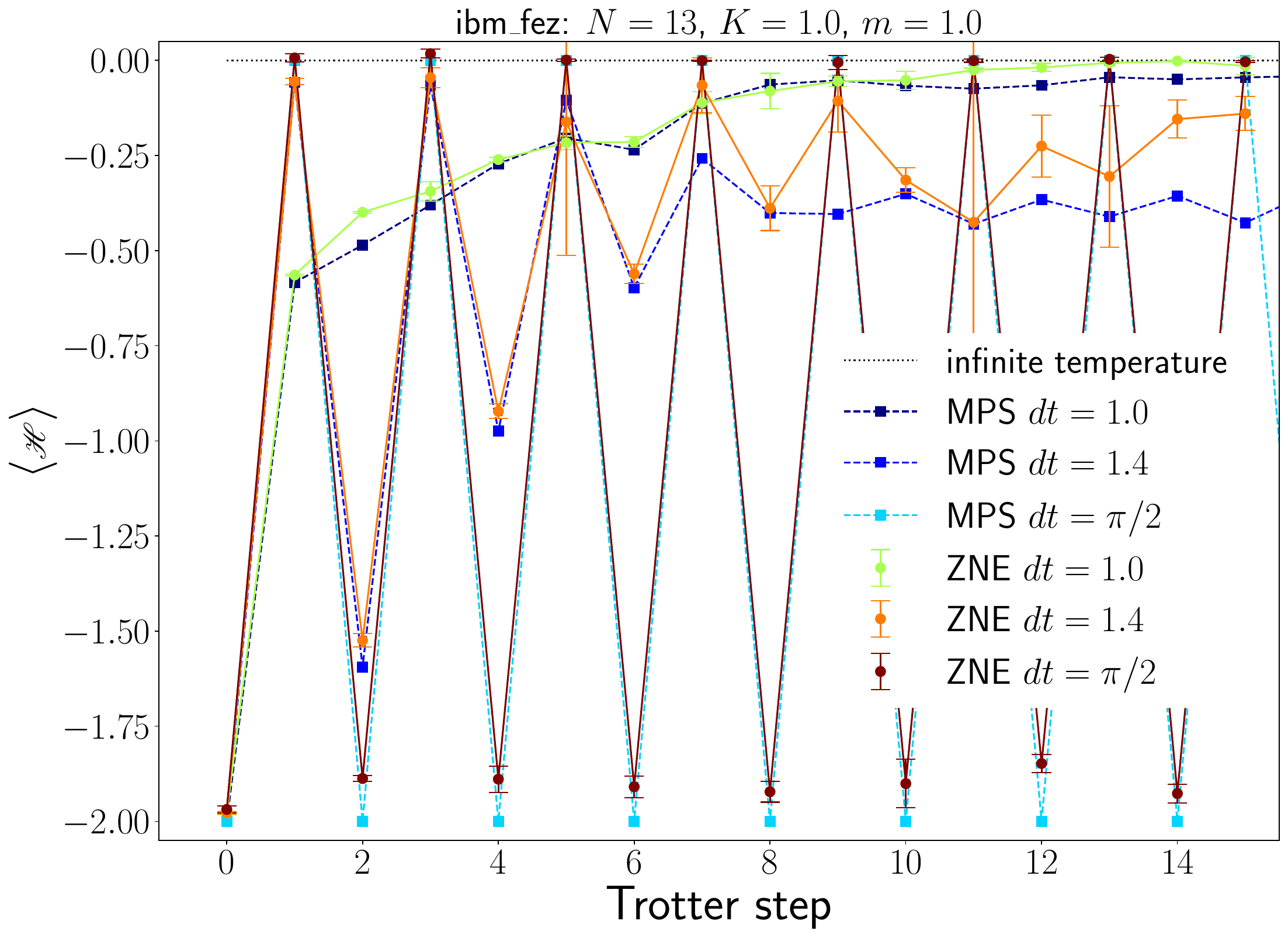}
  \caption{
  \label{fig:ZNE_N13}
   Quantum simulation results of Trotter-time step dependence of Floquet prethermalization using ibm\_fez. The system size is $N=13$ ($38$ qubits on the real device). The gauge parameter is $K=1.0$, and the fermion mass is $m=1.0$. ``MPS" shows the results of the classical MPS calculations, and ``ZNE" shows the mitigated results by ZNE. Lines are just for eye guides.
  }
\end{figure}

The simulation results of ZNE are shown in Fig.~\ref{fig:ZNE_N13}. (See also the raw data in Figs.~\ref{fig:gauss_N13} and~\ref{fig:pi2_N13}.) The ZNE results quantitatively agree with the MPS results in small Trotter steps. In particular, quantum simulations can nicely reproduce the MPS results up to $10$ Trotter steps for $dt=1.0$. This demonstrates that the current quantum device has the power to simulate physics-motivated problems, that is, thermalization dynamics and the very early stage of prethermal plateau. On the other hand, the results of $dt=1.4$ are not as good as those of $dt=1.0$, particularly, in $7$-$10$ Trotter steps. 
The origin of this difference is discussed in detail in Sec.~\ref{sec:discussion}.

ZNE seems to work remarkably well. 
Nevertheless, we cannot say that ZNE is the best one.
The computational costs of ZNE are very large.
In fact, we can simulate only up to $15$ Trotter steps if we enable both of the Pauli twirling and ZNE. In the following subsections, we employ two other methods that are less computationally demanding, and may be useful for simulating systems with a large number of qubits.

\subsection{mitigation by division: the Gauss law operator}
\begin{figure}[t]
  \centering
  \includegraphics[width=.48\textwidth]
  {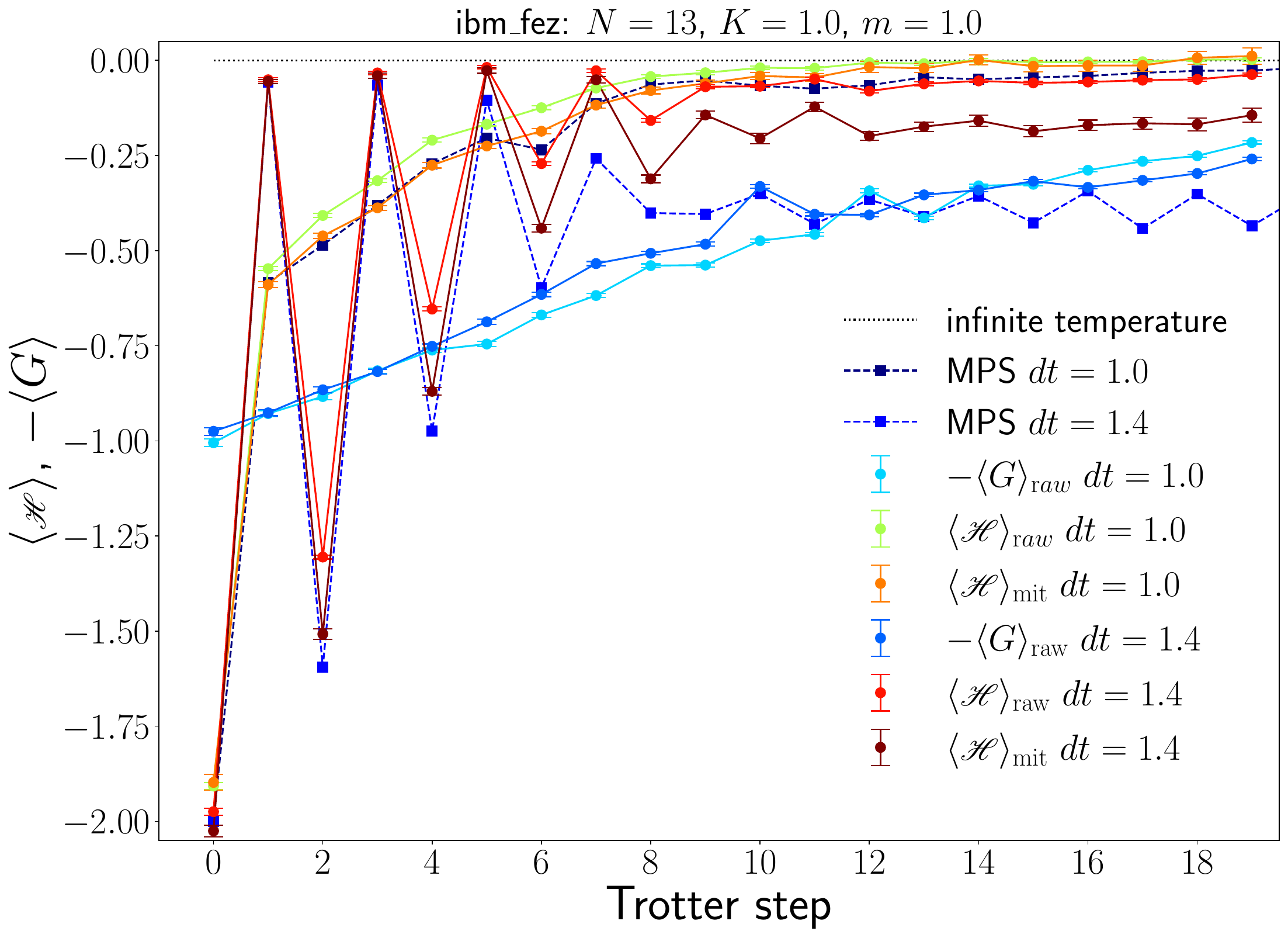}
  \caption{
  \label{fig:gauss_N13}
    Quantum simulation results of Trotter-time step dependence of Floquet prethermalization using ibm\_fez. The system size is $N=13$ ($38$ qubits on the real device). The gauge parameter is $K=1.0$, and the fermion mass is $m=1.0$. ``MPS" shows the results of the classical MPS calculations. $\langle G \rangle_{\rm raw}$ is the measured value of the Gauss law operator, $\langle \mathscr{H} \rangle_{\rm raw}$ is the measured value of the local Hamiltonian, and $\langle \mathscr{H} \rangle_{\rm mit}$ is the mitigated result by a division by $\langle G \rangle_{\rm raw}$.  Lines are just for eye guides. 
  }
\end{figure}

To correct amplitude decay due to decoherence, we need to estimate a rescaling factor. 
Lattice gauge theories have a very nice property useful for estimating the factor. 
The Gauss law constraint is an exact relationship in gauge theories and its expectation value is protected by local gauge symmetry.
The mitigation by it is universally applicable to any gauge theory.

If the initial state satisfies the Gauss law constraint $\langle G(x)\rangle=1$, it must not be violated during time evolution, but it is violated in noisy simulation.
As shown in Fig.~\ref{fig:gauss_N13}, the expectation value of the Gauss law operator decreases as the Trotter step increases. 
We identify the expectation value of the Gauss law operator as a damping factor, and perform error mitigation as
\begin{equation}
\langle \mathscr{H}(x_c) \rangle_{\rm mit}=
\frac{\langle \mathscr{H}(x_c) \rangle_{\rm raw}}{\langle G(x_c) \rangle_{\rm raw}}  .
\label{eqGmit}
\end{equation}
Here, the data obtained by enabling only TREX and Pauli twirling of two qubit gates are referred to as raw data.
The raw data and the mitigated results are shown in Fig.~\ref{fig:gauss_N13}.
The error mitigation works and quantum simulations agree with the MPS results for small Trotter steps. 
However, we see that mitigation by the Gauss law cannot capture the dip at $6$ Trotter steps for $dt=1.0$ that ZNE can. Again, although quantum simulations quantitatively agree with the MPS results up to $10$ Trotter steps for $dt=1.0$, the results of $dt=1.4$ are not as good as those of $dt=1.0$ in $6$-$10$ Trotter steps. 

While the local Hamiltonian $\mathscr{H}(x_c)$ acts on two qubits (two fermions), the Gauss law operator $G(x_c)$ acts on four qubits (two fermions and two gauge links). We expected that the difference of operator sizes may not affect after many gates are applied, but the results cannot reproduce the amplitudes in prethermal plateau. This might imply that it is important making operator sizes same in the numerator and denominator of Eq.~\eqref{eqGmit}.

\subsection{mitigation by division: \texorpdfstring{$dt=\frac{\pi}{2}$}{pi/2} circuit}
\begin{figure}[t]
  \centering
  \includegraphics[width=.48\textwidth]
  {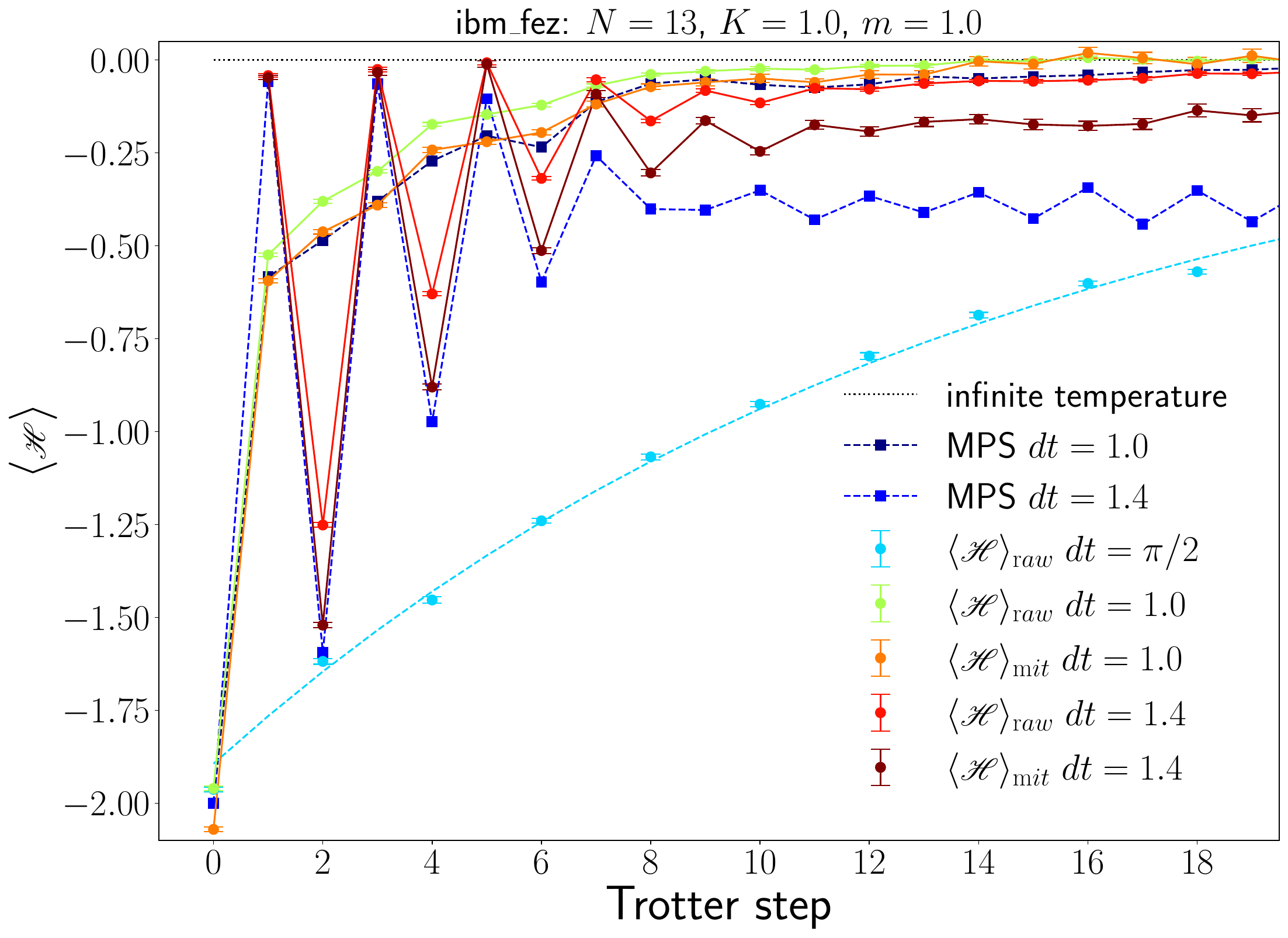}
  \caption{
  \label{fig:pi2_N13}
    Quantum simulation results of Trotter-time step dependence of Floquet prethermalization using ibm\_fez. The system size is $N=13$ ($38$ qubits on the real device). The gauge parameter is $K=1.0$, and the fermion mass is $m=1.0$. ``MPS" shows the results of the classical MPS calculations. $\langle \mathscr{H} \rangle_{\rm raw}$ is the measured value and $\langle \mathscr{H} \rangle_{\rm mit}$ is the mitigated result by a division by the fitting result of $dt=\pi/2$.  Lines are just for eye guides, while the curve is the result of exponential fitting of $\pi/2$ that is used in error mitigation of $dt=1.0$, and $1.4$.
  }
\end{figure}
\begin{figure}[ht]
  \centering
  \includegraphics[width=.48\textwidth]
  {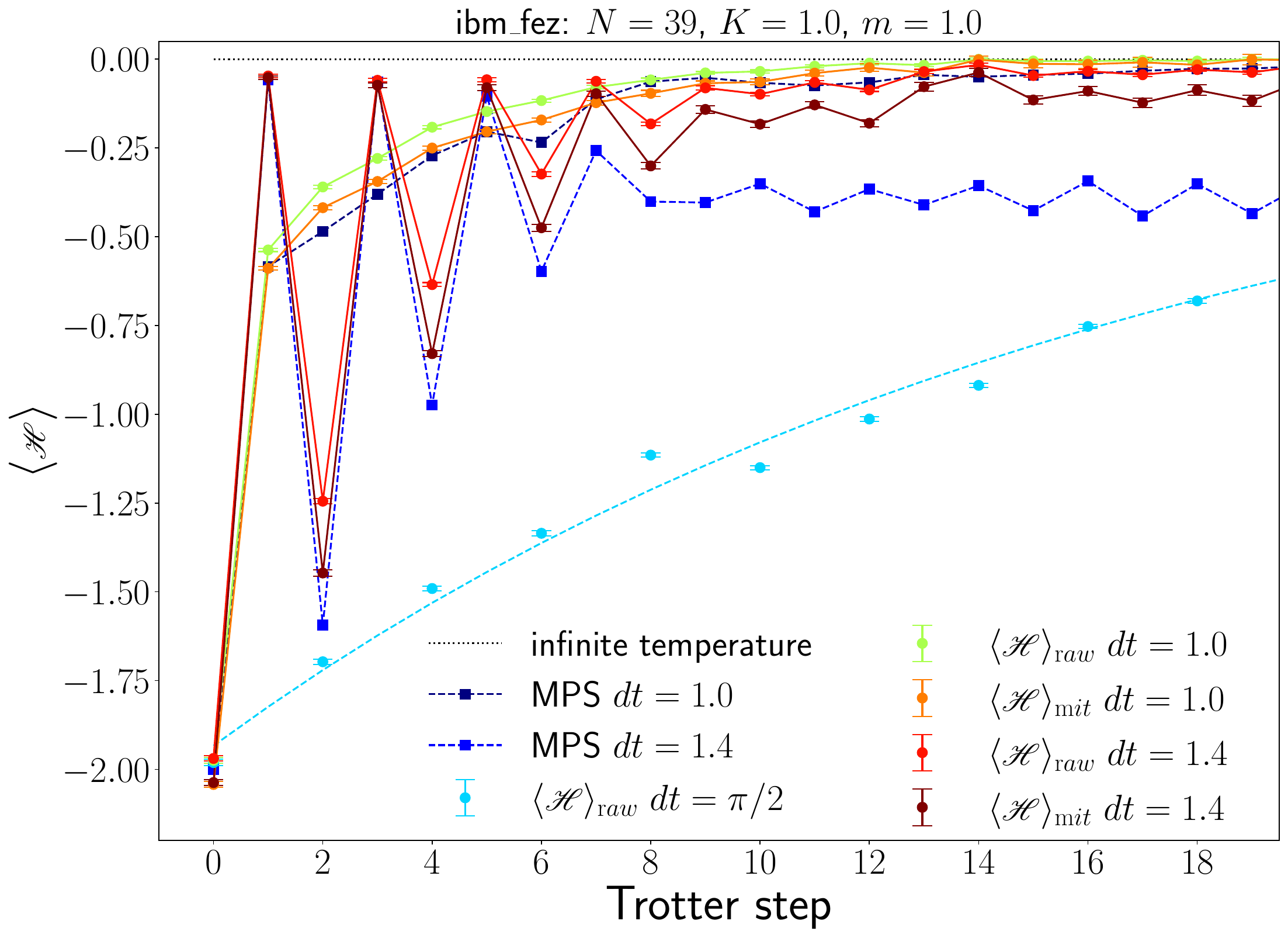}
  \caption{
  \label{fig:pi2_N39}
    Quantum simulation results of Trotter-time step dependence of Floquet prethermalization using ibm\_fez. The system size is $N=39$ ($116$ qubits on the real device). The gauge parameter is $K=1.0$, and the fermion mass is $m=1.0$. ``MPS" shows the results of the classical MPS calculations. $\langle \mathscr{H} \rangle_{\rm raw}$ is the measured value and $\langle \mathscr{H} \rangle_{\rm mit}$ is the mitigated result by a division by the fitting result of $dt=\pi/2$.  Lines are just for eye guides, while the curve is the result of exponential fitting of $\pi/2$ that is used in error mitigation of $dt=1.0$, and $1.4$.
  }
\end{figure}

To estimate a rescaling factor using the same operator that we will measure, we employ the error mitigation used in Ref.~\cite{Shinjo:2024vci}. 
We simulate the circuit with special parameters such that the circuit is exactly solvable if there is no error. In our case, if we set $dt=\frac{\pi}{2}$, the expectation value of the local Hamiltonian takes $-2\rightarrow0\rightarrow-2\rightarrow0\cdots$. 
In Fig.~\ref{fig:pi2_N13}, the actual data exponentially decrease as the Trotter step $N_t$ increases.
Using even $N_t$, we fit the results of the noisy simulations by assuming a form of 
\begin{equation}
\langle \mathscr{H}_{\frac{\pi}{2}}(x_c) \rangle_{\rm raw} =-A e^{-\lambda N_t}, 
\end{equation}
where $A$ and $\lambda$ are fitting parameters.
Using the fitting results, we perform error mitigation of $dt\neq\frac{\pi}{2}$ as
\begin{equation}
\langle \mathscr{H}(x_c) \rangle_{\rm mit} = \frac{2}{A} e^{\lambda N_t}\langle \mathscr{H}(x_c) \rangle_{\rm raw}  .
\end{equation}
Here, the data obtained by enabling only TREX and Pauli twirling of two qubit gates are referred to as raw data. The results of error mitigation are shown with the raw data in Fig.~\ref{fig:pi2_N13}. In fact, the results of error mitigation are very similar to those based on the Gauss law operator. Although the operator size is the same in this error mitigation method, the effective circuit volume may largely differ in thermalized and trivial circuits. This may be the reason why we cannot reproduce prethermal plateau in this error mitigation method. We may need a solvable circuit such that the growth of operator sizes is the same as the circuit that we would like to simulate, but it is in general hard to prepare such a circuit. This strongly restricts the number of Trotter steps that we can simulate correctly, and we can rely on ZNE if the available Trotter step is short.

Finally, we preform a large-scale simulation that uses more than $100$ qubits.
The system size is $N=39$, for which we use 116 of the 156 qubits of ibm\_fez. 
We show the results of quantum simulations in Fig.~\ref{fig:pi2_N39}. 
We used the error mitigation based on division. This may not be surprising since $N=13$ is large enough to compute the observable in a large system as seen in MPS simulations, but we see that the simulations that use more than $100$ qubits even quantitatively agree with the MPS results. This demonstrates the capability of IBM's superconducting quantum devices for studying quantum many-body dynamics in the era of quantum utility scale.

\subsection{Discussion}
\label{sec:discussion}
\begin{figure*}[t]
  \centering
  \includegraphics[width=.93\textwidth]
  {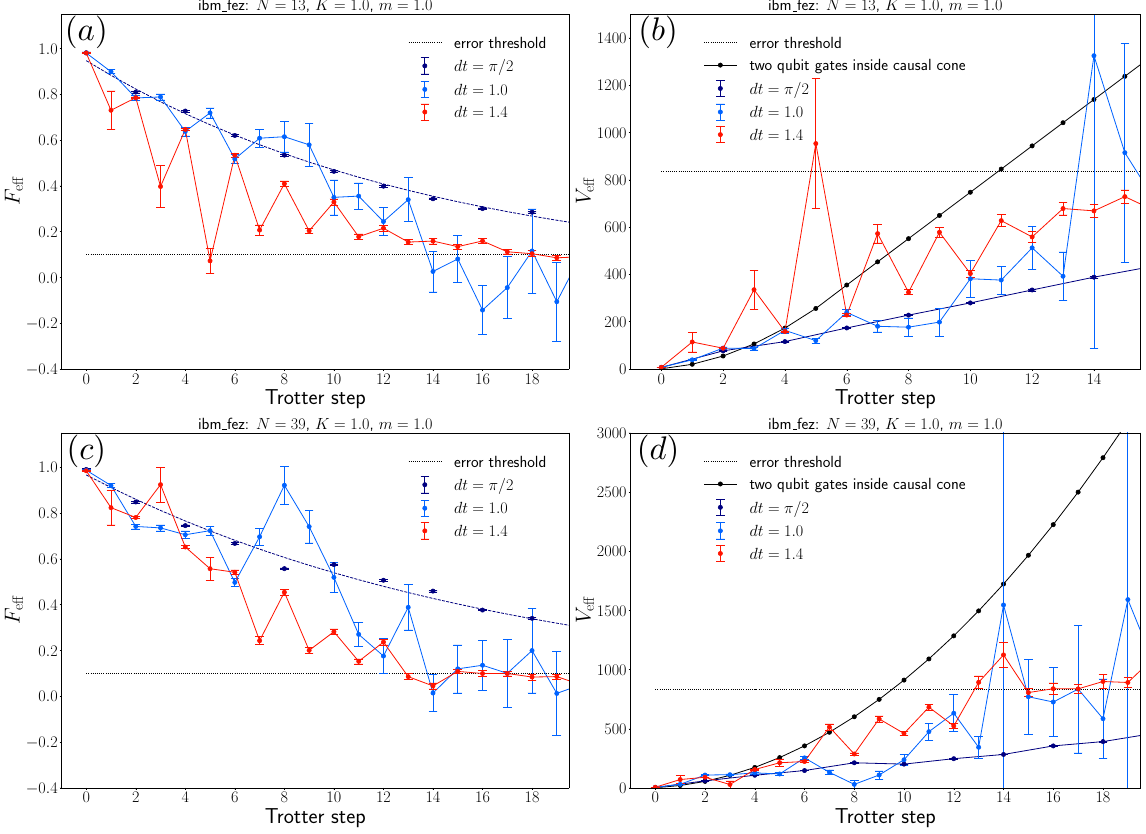}
  \caption{
  \label{fig:pi2_N13_fid}
    Effective fidelity $F_{\rm eff}$ of quantum circuits simulated using ibm\_fez for $N=13$ (a) and for $N=39$ (c), and corresponding circuit volume $V_{\rm eff}$ for $N=13$ (b) and for $N=39$ (d). The gauge parameter is $K=1.0$, and the fermion mass is $m=1.0$. We estimate $F_{\rm eff}$ by dividing $\langle \mathscr{H} \rangle_{\rm raw}$ by the corresponding ``MPS" results in Fig.~\ref{fig:pi2_N13}.  Lines are just for eye guides, while the curve is the result of exponential fitting of $\pi/2$. $V_{\rm eff}$ is related to $F_{\rm eff}$ as $F_{\rm eff}=(1-p)^{V_{\rm eff}}$, and we used $p=2.75\times 10^{-3}$, which is the median CZ error of ibm\_fez when we executed the circuits. A typical value of ``error threshold" $F_{\rm eff}=0.1$ in (a) and (c) corresponds to $V_{\rm eff}=836$ in (b) and (d). We show only up to $15$ Trotter steps in (b), since the fidelity in (a) becomes negative due to large noises above $15$ Trotter steps. For comparison, we show the number of two-qubit gates inside causal cones in (b) and (d). }
\end{figure*}

Assuming the global depolarizing channel, the experimental results before error mitigation $\langle \hat{O}\rangle_{\rm raw}$ are related with ideal results without noises $\langle \hat{O}\rangle_{\rm ideal}$ as~\cite{Kechedzhi:2023swt}
\begin{equation}
    \langle \hat{O}\rangle_{\rm raw} =F_{\rm eff} \langle \hat{O}\rangle_{\rm ideal} +\frac{1-F_{\rm eff}}{2^{3N-1}}\Tr{\hat{O}},
\end{equation}
with $F_{\rm eff}=(1-p)^{V_{\rm eff}}$, where $F_{\rm eff}$, $V_{\rm eff}$, $p$ are the effective fidelity, circuit volume, and typical value of gate errors, respectively.
We here estimate $F_{\rm eff}$ and $V_{\rm eff}$ to deepen understanding of the quantum simulation results.
We use the raw data $\langle \mathscr{H}(x_c) \rangle_{\rm raw}$, and the MPS results as $\langle \mathscr{H}(x_c) \rangle_{\rm ideal}$. We set $p=2.75\times 10^{-3}$, which is the median CZ error of ibm\_fez in executing the circuits.
$F_{\rm eff}$ is related with the experimentally achievable signal-to-noise ratio of the observable. If we set $F_{\rm eff}=0.1$ as a typical value for the fidelity threshold, this is equal to $V_{\rm eff}=836$ in terms of the CZ gates.

We show the estimated values of the effective fidelity $F_{\rm eff}$ and circuit volume $V_{\rm eff}$ in Fig.~\ref{fig:pi2_N13_fid}. For comparison, we show the results of $N=13$ and $N=39$.
In both system sizes, $F_{\rm eff}$ of $dt=1.0$ is close to that of $dt=\frac{\pi}{2}$ up to $10$ Trotter steps. 
This fact supports our observation that the mitigation by the $dt=\frac{\pi}{2}$ circuit works well for $dt=1.0$.
On the other hand, $F_{\rm eff}$ of $dt=1.4$ decays more rapidly, particularly in the $7$ -$10$ Trotter steps. The high effective fidelity of $dt=1.0$ (that is, the smallness of the effective causal cone seen in $V_{\rm eff}$) is considered to be the reason why we can simulate larger Trotter steps for $dt=1.0$ using ZNE.

The effective circuit volume $V_{\rm eff}$ is interpreted as the effective number of entangling gates contributing to expectation values.
The estimated values (except for the peak at $5$ Trotter steps for $dt=1.4$) are much below the threshold value $V_{\rm eff}=836$ up to $10$ Trotter steps.
This implies that the simulation output contains reliable signals in this range.
$V_{\rm eff}$ of $dt=1.0$ increases linearly with Trotter steps up to $10$ Trotter steps. 
$V_{\rm eff}$ of $dt=1.4$ exhibits similar linearity in a broader range, although it has a more complex structure, as expected from the oscillatory behavior of the observable. 
Physically, such a time interval corresponds to thermalization dynamics until the onset of a prethermal plateau. Importantly for us, $V_{\rm eff}$ is much smaller than the number of two-qubit gates inside causal cones.
This implies that, at least in our system of the ${\bf Z}_2$ lattice gauge theory, thermalization dynamics can be simulated more easily than expected from the two-qubit gate counting. (Unfortunately, this may imply that it can be simulated easily in classical simulations as well.)

\section{Summary and outlook}
\label{sec:summary}

We have simulated the Floquet time evolution of the one-dimensional ${\bf Z}_2$ lattice gauge theory on a superconducting quantum computer.
Although current quantum computers have a strong limitation on circuit depth, it can be used for simulating short-time physics, such as the emergence of Floquet prethermal plateau, by implementing error mitigation. This work demonstrates that current quantum computing has potential for application to high-energy physics problems. We should challenge quantum simulations of more general cases, for example, two- or three-dimensional lattice gauge theories~\cite{Yamamoto:2020eqi,Gustafson:2020yfe,Zohar:2021nyc,Lumia:2021tpu,Zache:2023dko,Hayata:2023puo,Hayata:2023bgh}. 
Simulating Floquet prethermalization in higher-dimensional systems would be one of the promising directions to show the quantum advantage. 
Also, once thermalized states are made by the simulation, it would be interesting to use them to study the dynamical properties of finite-temperature systems. We leave the detailed studies of those problems in the future.

\begin{acknowledgments}
The authors thank Yuta Kikuchi for his kind support.
The classical simulations were performed using cluster computers at iTHEMS in RIKEN and the quantum simulations were performed using IBM’s superconducting quantum computer on the cloud in IBM Quantum services, supported by UTokyo Quantum Initiative.
The views expressed are those of the authors, and do not reflect the official policy or position of IBM or the IBM Quantum team. 
This work was supported by JSPS KAKENHI Grant No.~19K03841, 22K03520, 24K00630.
\end{acknowledgments}

\bibliographystyle{apsrev4-2}
\bibliography{paper}

\end{document}